\documentclass[useAMS,usenatbib]{mn2e}
\usepackage{txfonts}
\usepackage{graphicx,times}
\usepackage{natbib}
\usepackage{amssymb}
\usepackage{longtable}
\usepackage{subfigure}

\title[Ages and Masses of 0.64 million RGB stars from the LAMOST Galactic Spectroscopic Survey]{Ages and Masses of 0.64 million Red Giant Branch stars from the LAMOST Galactic Spectroscopic Survey}
\author[Yaqian Wu]{Yaqian Wu$^{1,2}$\thanks{E-mail:
wuyaqian@nao.cas.cn; gzhao@nao.cas.cn}, Maosheng Xiang$^{1,3}$\thanks{E-mail: msxiang@nao.cas.cn}, Gang Zhao$^{1}$, Shaolan Bi$^{2}$, Xiaowei Liu$^{4}$, Jianrong Shi$^{1}$
 \newauthor  Yang Huang$^{4}$, Haibo Yuan$^{2}$, Chun Wang$^{5}$, Bingqiu Chen$^{4}$, Zhiying Huo$^{1}$, Juanjuan Ren$^{1}$
 \newauthor Zhijia Tian$^{4}$, Kang Liu$^{2}$, Xianfei Zhang$^{2}$, Yaguang Li$^{2}$, Jinghua Zhang$^{2}$ \\
$^{1}$Key Laboratory of Optical Astronomy, National Astronomical Observatories, Chinese Academy of Sciences,
             Beijing 100012, P.\ R.\ China;\\
$^{2}$Department of Astronomy, Beijing Normal University,
             Beijing 100875, P.\ R.\ China;\\
$^{3}$Max-Planck Institute for Astronomy, K{\"o}nigstuhl, D-69117,
             Heidelberg, Germany;\\
$^{4}$South West Institute For Astronomy Research, Yunnan University
             Kunming 650500, P.\ R.\ China;\\
$^{5}$Department of Astronomy, Peking University,
             Beijing 100871, P.\ R.\ China;\\
             }

\begin{document}


\pagerange{\pageref{firstpage}--\pageref{lastpage}} \pubyear{2017}

\maketitle

\label{firstpage}

\begin{abstract}
We present a catalog of stellar age and mass estimates for a sample of 640\,986 red giant branch (RGB) stars of the Galactic disk from the LAMOST Galactic Spectroscopic Survey (DR4). The RGB stars are distinguished from the red clump stars utilizing period spacing derived from the spectra with a machine learning method based on kernel principal component analysis (KPCA). Cross-validation suggests our method is capable of distinguishing RC from RGB stars with only 2 per cent contamination rate for stars with signal-to-noise ratio (SNR) higher than 50. The age and mass of these RGB stars are determined from their LAMOST spectra with KPCA method by taking the LAMOST - $Kepler$ giant stars having asteroseismic parameters and the LAMOST-TGAS sub-giant stars based on isochrones as training sets. Examinations suggest that the age and mass estimates of our RGB sample stars with SNR $>$ 30 have a median error of 30 per cent and 10 per cent, respectively. Stellar ages are found to exhibit positive vertical and negative radial gradients across the disk, and the age structure of the disk is strongly flared across the whole disk of $6<R<13$\,kpc. The data set demonstrates good correlations among stellar age, [Fe/H] and [$\alpha$/Fe]. There are two separate sequences in the [Fe/H] -- [$\alpha$/Fe] plane: a high--$\alpha$ sequence with stars older than $\sim$\,8\,Gyr and a low--$\alpha$ sequence composed of stars with ages covering the whole range of possible ages of stars. We also examine relations between age and kinematic parameters derived from the Gaia DR2 parallax and proper motions. Both the median value and dispersion of the orbital eccentricity are found to increase with age.
The vertical angular momentum is found to fairly smoothly decrease with age from 2 to 12\,Gyr, with a rate of about $-$50\,kpc\,km\,s$^{-1}$\,Gyr$^{-1}$. A full table of the catalog is public available online.
\end{abstract}


\section{Introduction}

Understanding the stellar population and assemblage history of the Milky Way requires detailed information for a large number of stars in full dimensionality, including 3D positions, 3D velocities, mass, age, and elemental abundances. Thanks to the implement of the Gaia space mission \citep{Gaia16,Gaia18} and a few ground-based large-scale spectroscopic surveys, such as the the LAMOST Experiment for Galactic Understanding and Exploration \citep[LEGUE;][]{Deng12,Zhao12}, the Apache Point Observatory Galactic Evolution Experiment \citep[APOGEE;][]{Majewski10} and GALAH \citep{Freeman12,De15}, such stellar data set now can be achieved.

Among those stellar information (parameters), the age can hardly be observed or measured directly but usually have to be inferred indirectly relying on stellar models \citep{Soderblom10}. A widely adopted method is to match the stellar atmospheric parameters with stellar isochrones in the H-R diagram. The method delivers robust stellar ages for millions of stars \citep[e.g.][]{Xiang17a,Mints17,Sanders18}. However, the method only work well for main-sequence turn-off and sub-giant stars, but cannot give precise ages for red giant stars because stellar isochrones of different ages are tightly crowded together in the red giant phases. Red giant stars are important tracers of the Galactic structure due to their bright luminosities and large numbers in the Galaxy. Delivering reliable age estimates for a large sample of red giant stars is timely for the study of Galactic structure and archaeology.

Since the age of a low-mass red giant star is dominated by the life time of its main-sequence evolutionary phase, one can obtain good age estimate from stellar evolutionary tracks if the stellar mass is known accurately \citep{Ness2016,Martig16,Wu18}. With this approach,
Martig et al. (2016) obtained ages for 1475 APOKASC stars \citep{Pinsonneault14} that have mass estimates based on asteroseismology of the $Kepler$ data.
The APOKASC sample is further adopted as a training data-set to deliver mass and age estimates from the APOGEE and LAMOST spectra with either a data-driven approach \citep{Ness2016, Ho17, Ting18} or an empirical relation with C and N abundances \citep{Martig16, Ness2016, Sanders18}.

With this approach, Wu et al. (2018) obtained precise ages of 6940 red giant branch (RGB) stars in the LAMOST-$Kepler$ project (De Cat et al. 2015, Ren et al. 2016, Zong et al. 2018), whose masses are determined with asteroseismic scaling relation using the seismic parameters of Yu et al. (2018). Typical uncertainty of the mass estimates is a few per cent, and that of the age estimates is $\sim$ 20 per cent. Wu et al. (2018) also investigated the feasibility of deducing age and mass directly from the LAMOST spectra with a machine learning method based on kernel based principal component analysis (KPCA) by taking the LAMOST-$Kepler$ sample stars as a training data set, and show that the stellar age thus derived achieve a precision of $\sim$ 24 per cent.

In this paper, we apply our method in Wu et al. (2018) to the LAMOST DR4 spectra database, and present a catalog of mass and age estimates for RGB stars in the LAMOST forth data release (DR4). A challenge in our work is to distinguish RGB stars from the red clump (RC) stars in the LAMOST database. In the $T_{\rm eff}$ -- {\rm log}\,$g$ diagram, the RGB and RC stars have overlaps due to metallicity effect. Since the metallicity determinations from the spectra have considerable errors, a simple cut in the $T_{\rm eff}$ -- {\rm log}\,$g$ to distinguish RC and RGB stars \citep[e.g.][]{Bovy14,Huang15} may lead to considerable contaminations or undesired patterns.  On the other hand, it has been suggested that the C and N abundances in the photosphere of a low-mass star change continuously as the star ascending the red giant branch due to non-canonical extra mixing processes (e.g. Charbonnel 1995, Charbonnel et al. 1998, Martell et al. 2008, Masseron et al. 2017, Hawkins et al. 2018). It is therefore expected that RC stars have different photospheric C and N abundances respect to their RGB progenitors, and thus can be distinguished from RGB stars using spectra (e.g. Hawkins et al. 2017, Ting et al. 2018).

In this work, we estimate the period spacing $\Delta P$ from the spectra with a machine learning method by taking the LAMOST-$Kepler$ sample as training dataset, and classify the RGB and RC stars with $\Delta P$ and log\,$g$. Our method is found to have a contamination rate of only 2 per cent to select RGB stars given high ($>$ 50) spectral signal-to-noise ratio (SNR). With this method, we select 866\,315 RGB spectra from LAMOST DR4. The sample stars are found to have a median error of 10 per cent in mass and 30 per cent in age estimates for stars with SNR $>$ 30. For 0.20 million stars that have SNR higher than 50,  the age estimates have random errors even smaller than 20 per cent. With
this large RGB sample, we investigate the distribution of
stellar ages in the disk $R$ -- $Z$ plane, the relations between stellar
age, metallicity and abundances, as well as relations between age and kinematic
parameters derived from the Gaia DR2 parallax and proper
motions.

The paper is organized as follows. Section\,2 introduces the LAMOST data and how to select the LAMOST RGB star sample. Section\,3 describes the mass and age estimation of the LAMOST RGB stars. Statistical properties of the RGB samples is presented in Section\,4, followed by conclusions in Section\,5.

\section{Select LAMOST RGB sample}

\subsection{LAMOST Data}
In the forth data release (DR4), the LAMOST Galactic Survey \citep{Deng12,Zhao06,Zhao12} released more than 7 million spectra observed with LAMOST \citep{Cui12} by June, 2016. Stellar parameters, including radial velocity $V_{\rm r}$, effective temperature $T_{\rm eff}$, surface gravity log\ $g$, and metallicity [Fe/${\rm H}$], are delivered from the spectra with the LAMOST Stellar Parameter pipeline \citep[LASP;][]{Wu11,Luo15}. The LASP stellar parameters are publicly available via the LAMOST official data releases \citep{Luo12,Luo15}.

Stellar parameters, including $V_{\rm r}$, $T_{\rm eff}$, log\,$g$, [Fe/H], interstellar reddening $E_{B-V}$, absolute magnitudes ${\rm M}_V$ and ${\rm M}_{K_{\rm s}}$, $\alpha$-element to metal (and iron) abundance ratio [$\alpha$/M] and [$\alpha$/Fe], as well as carbon and nitrogen abundance [C/H] and [N/H], have also been derived with the LAMOST Stellar Parameter Pipeline at Peking University \citep[LSP3;][]{Xiang15a, Xiang17b}, utilizing spectra processed with an independent flux calibration pipeline aiming to get better treatment with interstellar
extinction of the flux standard stars \citep{Xiang15b,Xiang17c}. Stellar parameters deduced with LSP3 for targets in the LAMOST Spectroscopic Survey of the Galactic Anticentre (LSS-GAC; Liu et al. 2014, Yuan et al. 2015), which is one of the major components of the LAMOST Galactic survey, as well as values of extinction, distance, and orbital parameters inferred using the LSP3 stellar parameters, are publicly released as the LSS-GAC valued-added catalogs \citep{Yuan15,Xiang17b}. An extended application of the LSP3 to the whole data set of LAMOST DR4 leads to a value-added catalog of LAMOST DR4, which contains $V_{\rm r}$, $T_{\rm eff}$, ${\rm M}_V$, log\,$g$,
[Fe/${\rm H}$], [$\alpha$/Fe], [C/H], [N/H], extinction and distance for 6\,597\,527 spectra of 4\,373\,824 stars. This value-add catalog of LAMOT DR4 has been also public available from the LAMOST official website\footnote{http://dr4.lamost.org/doc/vac}. Given a pixel spectral SNR higher than 50, stellar parameters yielded by LSP3 have typical precision of about 100\,K for $T_{\rm eff}$, 0.1\,dex for log\ $g$, 0.3 -- 0.4\,mag for $M_{V}$ and $M_{Ks}$, 0.1\,dex for [Fe/${\rm H}$], [C/H] and [N/H], and better than 0.05\,dex for [$\alpha$/M] and [$\alpha$/Fe] \citep{Xiang17c}. Utilizing the criteria of $T_{\rm eff}$ $<$ 5500\,K and log\ $g$ $<$ 3.8\,dex, we obtained more than 1.0 million LAMOST spectra of giant stars. This sample contains giant stars of all evolutionary stages (e.g. RGB, RC, AGB etc.), as the $T_{\rm eff}$ and log\,$g$ criteria themselves cannot clearly tell the evolutionary stages.

\subsection{Classification of RGB and RC}

In the HR diagram, both RGB and RC stars could appear at the same region (e.g. $T_{\rm eff}$ $\sim$ 4800\,K, log\ $g$ $\sim$ 2.4\,dex). However, in stellar physics, RGB and RC are two different evolutionary states. RGB stars are evolved stars that burn hydrogen in a shell around an inert helium core (Iben 1968) whereas RC stars burn with helium. Considering the fact that RC stars suffer from significant mass loss which is poorly constrained with the current data, and the impact of the mass loss on age estimation remained to be investigated in detail. In this work, we therefore focus on age estimation for RGB stars. In this section, we introduce how to select RGB stars from the above 1.0 million LAMOST red giant stellar spectra.

Asteroseismology has become the gold standard for distinguishing RGB and RC stars \citep{Montalban10,bedding11,Mosser11,Mosser12,Stello13,Pinsonneault14,Vrard16,Elsworth17}. The solar-like oscillations in red giant stars arise from near-surface convection and can have acoustic (p-mode) and gravity (g-mode) characteristics \citep{Chaplin13}. P modes are always associated with the stellar envelope with pressure as restoring force, while g modes are from the inner core with buoyancy as restoring force. The observed stellar pulsations of evolved stars are 'mixed modes' which include information of p modes and g modes. Given the same luminosity and radius, the core density of RC stars is lower than that of RGB stars, which causes a significantly stronger coupling between g- and p- modes and leads to larger period spacing (Bedding et al. 2013). Then we can distinguish RGB stars and RC stars accurately from large frequency spacing ($\Delta\nu$) and period spacing ($\Delta P$). However, currently it is hard to obtain accurate large frequency spacing and period spacing for large samples of giant stars utilizing asteroseismology. On the other hand, as introduced in Section 1, the RC stars have different C and N abundances respect to their RGB counterparts, as a consequence of non-canonical extra mixing processes in the RGB phase. It is therefore expected that there exists physical correlation between the period spacing and the photospheric C and N abundances for RGB and RC stars. So that it is plausible to build empirical (data-driven) model to infer the period spacing from stellar spectra by making full use of carbon and nitrogen features, and thus further clarify RGB and RC stars. Such a data-driven approach to classify RGB and RC has been shown to be practical and effective on APOGEE spectra by Hawkins et al. (2017), and has been successfully applied to APOGEE DR14 and LAMOST DR3 by Ting et al. (2018).
Here we estimate the $\Delta\nu$ and $\Delta P$ from the LAMOST spectra with a machine learning method based on KPCA, and then distinguish RGB and RC stars utilizing these two parameters.

Accurate large frequency spacing and period spacing for a few thousand stars have been derived from their power spectra with the \emph{Kepler} photometry \citep{Stello13,Mosser14,Vrard16,Elsworth17}. Among these results, we found that the sample of Vrard et al. (2016) is a most complete one with good accuracy. A cross-identification of the LAMOST-\emph{Kepler} stars with the sample of Vrard et al. (2016) yields 3973 LAMOST spectra of 3422 unique stars in common. Among them, 2705 stars have spectral signal-to-noise ratios (SNRs) higher than 50, including 807 RGB stars, 1623 RC stars and 235 secondary RC stars. Similar to method discussed in Wu et al. (2018), we divided this sample into two groups of equal number, one group is adopted as the training set to estimate $\Delta\nu$ and $\Delta P$ of the other, testing group of stars from the LAMOST spectra with a multivariate linear regression method. The regression method trains a relation between the parameters ($\Delta\nu$, $\Delta P$) and principal components derived from the spectra with the KPCA algorithm \citep{Sch98,Muller01,Xiang17b}. To avoid over-fitting, we have added a \textbf{$L2$} regularization term to the loss function (least squares) of the regression. Fig.\,1 plots the dispersion of the relative residuals of period spacing between results generated by the KPCA-based regression method from the spectra and the asteroseismic values for both the training and testing groups as a function of $N_{\rm PC}$. The dispersion for both the testing sample and the training sample gradually decreases with the increase of principal component (PC), and the trend becomes nearly flat at large number of PC. We therefore adopt 100 PCs as an optimal choice for the estimation of the $\Delta\nu$ and $\Delta P$ in this work.

\begin{figure}
\centering
\includegraphics[width=90mm, angle=0]{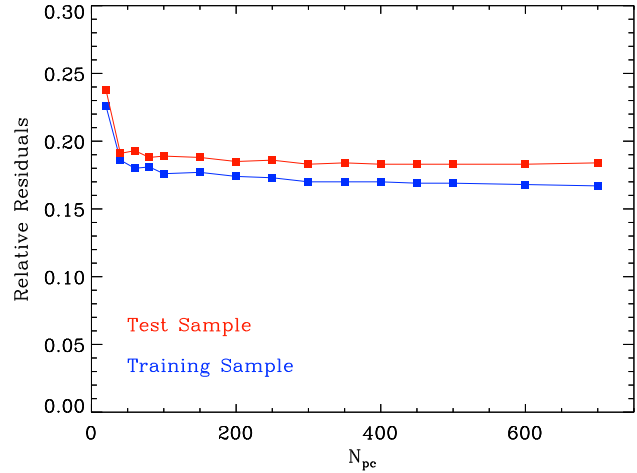}
\caption{The dispersion of the relative residuals of period spacing for both the training and test samples as a function of $N_{\rm PC}$. Red squares represent test sample, blue squares represent training sample. }
\label{Fig:1}
\end{figure}

The left panel of Fig.\,2 shows a comparison of period spacing deduced with the KPCA method and the seismic values for testing sample, with $N_{\rm PC}$ =100. Blue dots represent RGB stars, red dots represent RC stars and green dots represent secondary RC stars. For period spacing, the seismic values show two parts that are well separated, one is higher than 200s, which represents RC stars, and the other is lower than 100s which indicates RGB stars. It shows that our KPCA results from the spectra well reproduce the $\Delta P$ of RC stars. While although $\Delta P$ of the RGB stars are relatively poor estimated, most of them have values smaller than 200s, which are still distinguishable from that of the RC stars.

Since the LAMOST spectra contains information of stellar surface gravity, which is related to the large frequency spacing. We use the relationship between the surface gravity and the period spacing when we select RGB and RC from the spectra. The right panel of Fig.\,2 plots the distribution of period spacing and surface gravity with testing sample, these stellar period spacing are deduced from the LAMOST spectra.
It can be seen that the period spacing deduced from the KPCA could distinguish RGB stars and RC stars with the criteria of period spacing lower than 200s and larger than 200s. The period spacing of RGB stars are mostly lower than 200s and period spacing of RC stars are mostly larger than 200s. There are also a few stellar period spacing of RC stars mixing into RGB stars because their period spacing deduced from the spectra are lower than 200s.

\begin{figure*}
\centering
\includegraphics[width=150mm, angle=0]{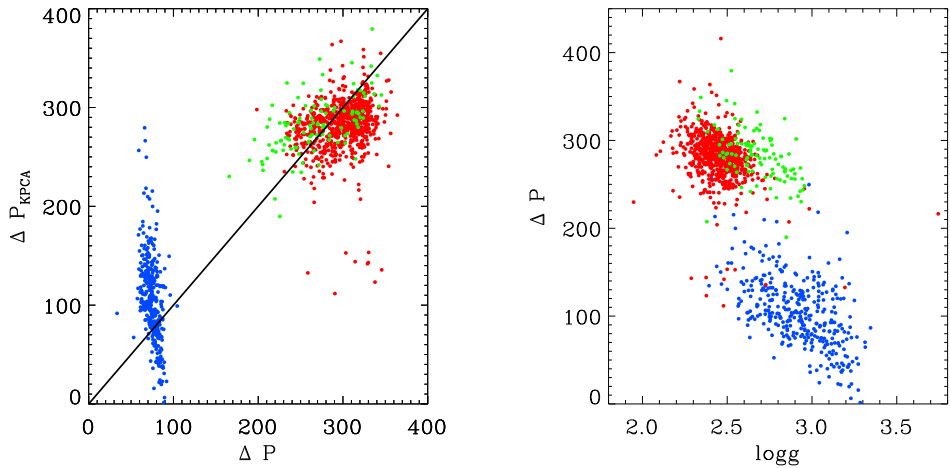}
\caption{$Left panel$: Comparison of period spacing $\Delta P$ estimated from the LAMOST spectra with the seismic values of Vrard et al. (2016) for the testing sample. The blue dots represent RGB stars, red dots represent RC stars and green dots represent secondary RC stars. $Right panel$: The period spacing and surface gravity diagram for the testing sample stars. The stellar period spacing are estimated from the LAMOST spectra. Colours have the same meanings as the left panel.}
\label{Fig:5}
\end{figure*}

The precision of period spacing deduced from the spectra are found to be sensitive to the spectral SNRs. Fig.\,3 plots the distribution of the estimated $\Delta P$ from the LAMOST spectra for the common stars with Vrard et al. (2016) in different spectral SNR bins: SNR $>$ 80, 40$\leq$SNR$\leq$80, SNR $<$40. The figure shows that in high SNR cases, RC and RGB stars are well separated in $\Delta P$, while in low SNR cases (e.g. $<20$), the separation in $\Delta P$ between RC and RGB stars are less clear due to large random error of the $\Delta P$ estimates. Nevertheless, there are still two clear peaks in the figure for ${\rm SNR} < 40$.

\begin{figure*}
\centering
\includegraphics[width=150mm, angle=0]{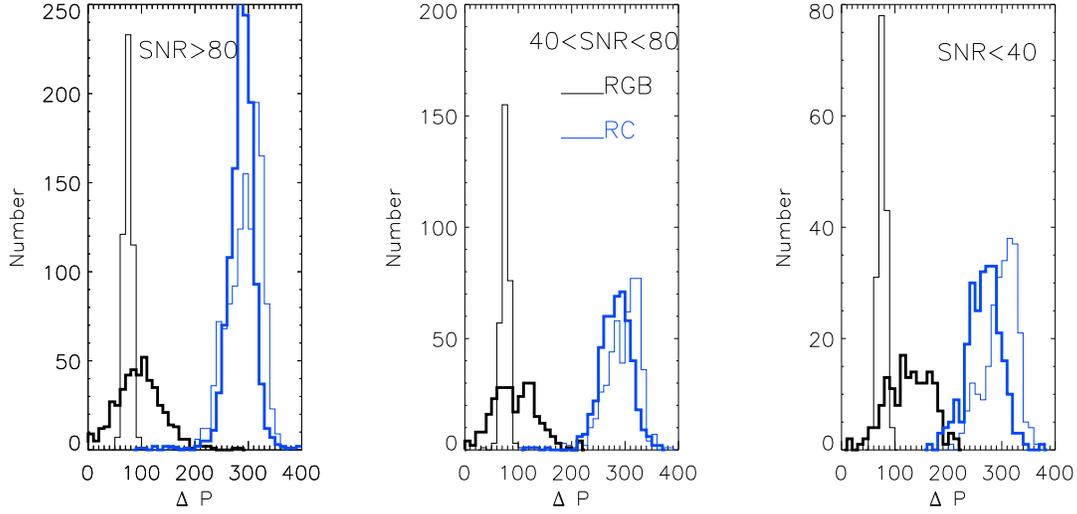}
\caption{Comparison of period spacing estimated from the spectra with seismic values for stars in different the SNR bins. Black and blue lines represent RGB and RC, respectively. Thin lines represent the classifications based on asteroseismology, while thick lines represent classifications based on spectra. }
\label{Fig:6}
\end{figure*}

We therefore adopt a maximum likelihood method to identify RGB stars and RC stars making use of $\Delta P$ and log\,$g$, and considering cases of different SNR bins (${\rm SNR}>80$, $40<{\rm SNR}<80$, ${\rm SNR}<40$) as well as different metallicity bins (${\rm [Fe/H]}<-0.4$, $-0.4<{\rm [Fe/H]}<0$, ${\rm [Fe/H]}<0$). Note that, there are only a few stars with SNR $< 40$ in the control sample, we thus opt not to divide those stars with lower SNRs into different metallicity bins.

In each SNR and [Fe/H] bin, the likelihood function is defined as
\begin{equation}
L_{i,j}=a_{i,j}\exp\frac{-(O_{i,j}-X_{i,j})^2}{2c_{i,j}^2},
\end{equation}
 where $O$ represents observational parameters, $X$ represents models, i represents surface gravity and period spacing, $i = 1$ represents surface gravity and $i =2$ represents period spacing; j represents the classification of RGB, RC and 2RC, $j=1,2,3$ represents RGB, RC and secondary RC, respectively. Then the probability of RGB stars, RC stars and secondary RC stars are
\begin{equation}
L_{j}=\frac{\prod_{i=1}^{2} L_{i,j}}{\sum_{j=1}^{3} L_{i,j}}.
\end{equation}
 We make a table for these parameters in Table\,1, character of $a$ represents the height of the Gaussian function, and character of $c$ represents the dispersion of each corresponding quantity.
 Utilizing the maximum likelihood, we calculate the probability of 1.0 million LAMOST stars. Stars with $L_{\rm rgb} > 0.5$ are selected as RGB stars, while the others are RC stars. In this work, we did not consider the pollution of AGB stars because of there are few AGB stars with seismic parameters.
 Finally, we note that a drawback of our maximum likelihood method is that the coefficients in Table\,1 are completely determined by the asteroseismic (training) sample, which are unlikely the optimal ones for the LAMOST data set. A further improvement should induce a Bayesian scheme considering priori of realistic populations of RGB and RC in different regions of the Galaxy.

\begin{table*}
\caption{Parameters adopted for the maximum likelihood method to distinguish RGB and RC stars in Equation\,1.}
\begin{tabular}{c}
 1),SNR$>$80, [Fe/H]$\leq$-0.4\\
\end{tabular}

\begin{tabular}{cccccc}
\hline
$a_{1,1}$=15.70 & $X_{1,1}$=2.76 & $c_{1,1}$=0.16 & $a_{2,1}$=26.68 & $X_{2,1}$=2.42 & $c_{2,1}$=0.09\\
$a_{1,2}$=15.49 & $X_{1,2}$=117.99 & $c_{1,2}$=45.34 & $a_{2,2}$=22.63 & $X_{2,2}$=292.79 & $c_{2,2}$=12.91\\
$a_{1,3}$=0.30 & $X_{1,3}$=2.71 & $c_{1,3}$=0.16 & $a_{2,3}$=0.30 & $X_{2,3}$=278.30 & $c_{2,3}$=26.20\\
\hline
\end{tabular}

\begin{tabular}{c}
 2),SNR$>$80, -0.4 $\leq$[Fe/H]$\leq$ 0\\
\end{tabular}

\begin{tabular}{cccccc}
\hline
$a_{1,1}$=26.09 & $X_{1,1}$=2.94 & $c_{1,1}$=0.20 & $a_{2,1}$=86.79 & $X_{2,1}$=2.47 & $c_{2,1}$=0.08\\
$a_{1,2}$=36.48 & $X_{1,2}$=92.78 & $c_{1,2}$=42.77 & $a_{2,2}$=86.86 & $X_{2,2}$=292.42 & $c_{2,2}$=16.94\\
$a_{1,3}$=9.60 & $X_{1,3}$=2.71 & $c_{1,3}$=0.16 & $a_{2,3}$=13.80 & $X_{2,3}$=278.30 & $c_{2,3}$=26.20\\
\hline
\end{tabular}

\begin{tabular}{c}
 3),SNR$>$80, [Fe/H]$>$ 0\\
\end{tabular}

\begin{tabular}{cccccc}
\hline
$a_{1,1}$=12.02 & $X_{1,1}$=3.10 & $c_{1,1}$=0.23 & $a_{2,1}$=25.29 & $X_{2,1}$=2.50 & $c_{2,1}$=0.14\\
$a_{1,2}$=21.66 & $X_{1,2}$=91.85 & $c_{1,2}$=34.49 & $a_{2,2}$=51.95 & $X_{2,2}$=277.30 & $c_{2,2}$=22.31\\
$a_{1,3}$=9.60 & $X_{1,3}$=2.71 & $c_{1,3}$=0.16 & $a_{2,3}$=13.80 & $X_{2,3}$=278.30 & $c_{2,3}$=26.20\\
\hline
\end{tabular}

\begin{tabular}{c}
 4),40 $\leq$SNR$\leq$ 80, [Fe/H]$\leq$-0.4\\
\end{tabular}

\begin{tabular}{cccccc}
\hline
$a_{1,1}$=18.24 & $X_{1,1}$=2.83 & $c_{1,1}$=0.31 & $a_{2,1}$=16.39 & $X_{2,1}$=2.37 & $c_{2,1}$=0.09\\
$a_{1,2}$=9.27 & $X_{1,2}$=110.88 & $c_{1,2}$=32.93 & $a_{2,2}$=16.41 & $X_{2,2}$=285.84 & $c_{2,2}$=24.38\\
$a_{1,3}$=0.20 & $X_{1,3}$=2.58 & $c_{1,3}$=0.20 & $a_{2,3}$=0.20 & $X_{2,3}$=275.07 & $c_{2,3}$=27.20\\
\hline
\end{tabular}

\begin{tabular}{c}
 5),40 $\leq$SNR$\leq$ 80, -0.4 $\leq$[Fe/H]$\leq$ 0\\
\end{tabular}

\begin{tabular}{cccccc}
\hline
$a_{1,1}$=21.95 & $X_{1,1}$=2.92 & $c_{1,1}$=0.18 & $a_{2,1}$=43.22 & $X_{2,1}$=2.44 & $c_{2,1}$=0.12\\
$a_{1,2}$=19.43 & $X_{1,2}$=87.27 & $c_{1,2}$=42.61 & $a_{2,2}$=40.43 & $X_{2,2}$=290.46 & $c_{2,2}$=25.98\\
$a_{1,3}$=3.80 & $X_{1,3}$=2.58 & $c_{1,3}$=0.20 & $a_{2,3}$=4.85 & $X_{2,3}$=275.07 & $c_{2,3}$=27.20\\
\hline
\end{tabular}

\begin{tabular}{c}
 6),40 $\leq$SNR$\leq$ 80, [Fe/H]$>$ 0\\
\end{tabular}

\begin{tabular}{cccccc}
\hline
$a_{1,1}$=12.26 & $X_{1,1}$=2.96 & $c_{1,1}$=0.24 & $a_{2,1}$=23.54 & $X_{2,1}$=2.48 & $c_{2,1}$=0.15\\
$a_{1,2}$=10.43 & $X_{1,2}$=94.93 & $c_{1,2}$=47.27 & $a_{2,2}$=27.68 & $X_{2,2}$=268.89 & $c_{2,2}$=22.99\\
$a_{1,3}$=3.80 & $X_{1,3}$=2.58 & $c_{1,3}$=0.20 & $a_{2,3}$=4.85 & $X_{2,3}$=275.07 & $c_{2,3}$=27.20\\
\hline
\end{tabular}

\begin{tabular}{c}
 7),SNR $<$ 40\\
\end{tabular}

\begin{tabular}{cccccc}
\hline
$a_{1,1}$=20.72 & $X_{1,1}$=2.90 & $c_{1,1}$=0.21 & $a_{2,1}$=49.34 & $X_{2,1}$=2.42 & $c_{2,1}$=0.17\\
$a_{1,2}$=17.39 & $X_{1,2}$=129.80 & $c_{1,2}$=45.69 & $a_{2,2}$=31.12 & $X_{2,2}$=267.63 & $c_{2,2}$=30.57\\
$a_{1,3}$=6.84 & $X_{1,3}$=2.65 & $c_{1,3}$=0.18 & $a_{2,3}$=5.30 & $X_{2,3}$=264.86 & $c_{2,3}$=36.25\\
\hline
\end{tabular}
\end{table*}

\subsection{The RGB Sample}

In order to assess the completeness and contamination rate of our method to select RGB stars, we use stars in the sample of Elsworth et al. (2017) that have LAMOST spectra but are not in our training sample as an independent validation. Elsworth et al. (2017) provides the classification of RGB, RC and secondary RC, but they do not provide their period spacing. The sample contains 3138 stars, including 1785 RGB, 1149 RC and 204 secondary RC. We calculate the likelihood of these stars to be red giant branch stars, and plot the seismic classification of this sample to be red giant branch stars and red clump stars or secondary red clump stars in Fig.\,4. The figure demonstrates that the distribution of likelihood of seismic classification for RGB stars and RC stars exhibits two sharp peaks, one with a value of zero and the other with a value of unity, indicating that most of RGB and RC stars can be distinguished very well, whereas there are a small number of stars with likelihood values around 0.4 to 0.6, which are hard to clarify whether they are RGB stars or RC stars. We define that stars with $L_{\rm rgb}$ $>$ 0.5 are RGB stars and then calculate their completeness and contamination rate. The completeness of this sample is found to be 94 per cent and the contamination rate is only 2 per cent. Note that most of stars have SNRs higher than 50, while we expect that the contaminate rate are higher than 2 per cent when we use this method to distinguish RGB stars and RC stars with lower SNRs. The completeness rate is higher and the contamination rate is lower than methods in literature based on stellar atmospheric parameters alone \citep{Bovy14,Huang15}, while it is comparable to the results of Hawkins et al. (2018) and Ting et al. (2018), which are based on similar philosophy to this work.

\begin{figure}
\centering
\includegraphics[width=90mm, angle=0]{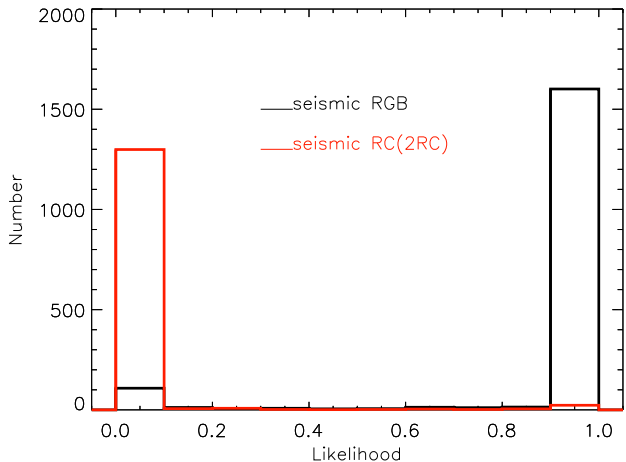}
\caption{The probability distribution of 3138 stars from the work of Elsworth et al. (2017), the black line represents the probability distribution of the seismology classified RGB stars, and the red line represents the probability distribution of the seismology classified RC and secondary RC stars. }
\label{Fig:7}
\end{figure}

We apply the method to the LAMOST DR4 spectra of giant stars. Utilizing the criteria of $L_{\rm rgb}$ $>$ 0.5, we obtain 866\,315 red giant stellar spectra, with 640\,986 unique stars. Their distribution of SNRs are shown in Fig.\,5, the median SNRs of these stars are 30. There are 0.20 million stars with SNRs greater than 50, 0.41 million stars with SNRs greater than 20. Considering the fact that the accuracy of period spacing varied from the SNRs, we select RGB stars with SNRs greater than 30 to study the properties of this RGB sample stars. We plot their H-R diagram in Fig.\,6, the color bar indicates their density. The effective temperatures of the sample stars cover the range of 3000 $-$ 5500\,K and the surface gravities vary from 0.5 to 3.8\,dex. From the figure, we can see that there is no obvious contaminations of red clumps, which causes a compact over-density at ${\rm log} g\sim2.4$ (see Fig.\,9 of Xiang et al. 2017b as an example). There are some stars located in the lower RGB phase. On the whole, the parameters space of RGB stars are completely covered with negligible pollution by red clump stars.

\begin{figure}
\centering
\includegraphics[width=90mm, angle=0]{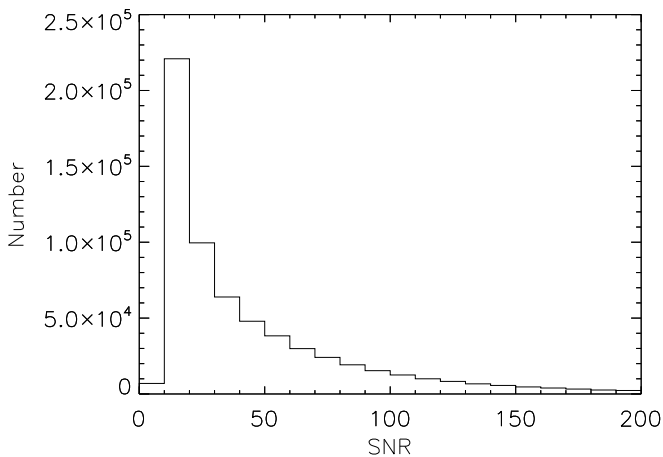}
\caption{SNR distribution of the LAMOST RGB stellar sample. }
\label{Fig:8}
\end{figure}

\begin{figure}
\centering
\includegraphics[width=90mm, angle=0]{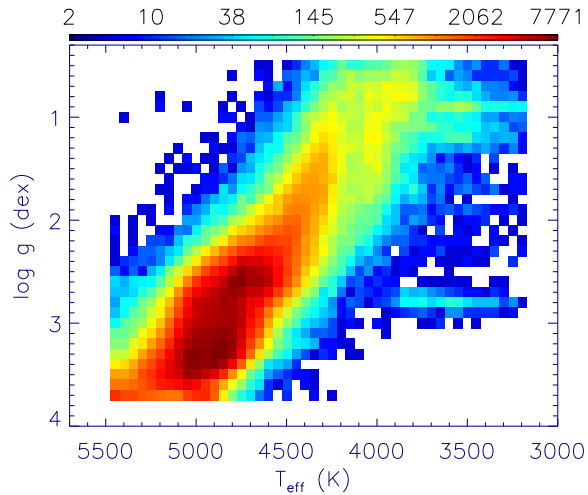}
\caption{The effective temperature -- surface gravity diagram for LAMOST RGB stars with SNR $>$ 30. The color bar represents the number of stars. }
\label{Fig:9}
\end{figure}

Fig.\,7 plots the density distribution of LAMOST RGB stars in the Galactic $X$ -- $Y$ and $X$ -- $Z$ plane for the RGB stars. Here, the $X$, $Y$ and $Z$ are computed assuming $R_{\odot}$ = 8.0\,kpc and $Z_{\odot}$ = 0\,kpc, and the recommended distances are from the value-added catalog of LAMOST DR4 (see Section 2). The figure shows that the most of the RGB sample stars are distributed in the range of $-$20\,kpc to 0\,kpc for $X$, $-$7\,kpc to 7\,kpc for $Y$ and $-$4\,kpc to 4\,kpc for $Z$. 46 per cent of the stars are located within 2kpc from the Sun.

\begin{figure*}
\centering
\includegraphics[width=150mm, angle=0]{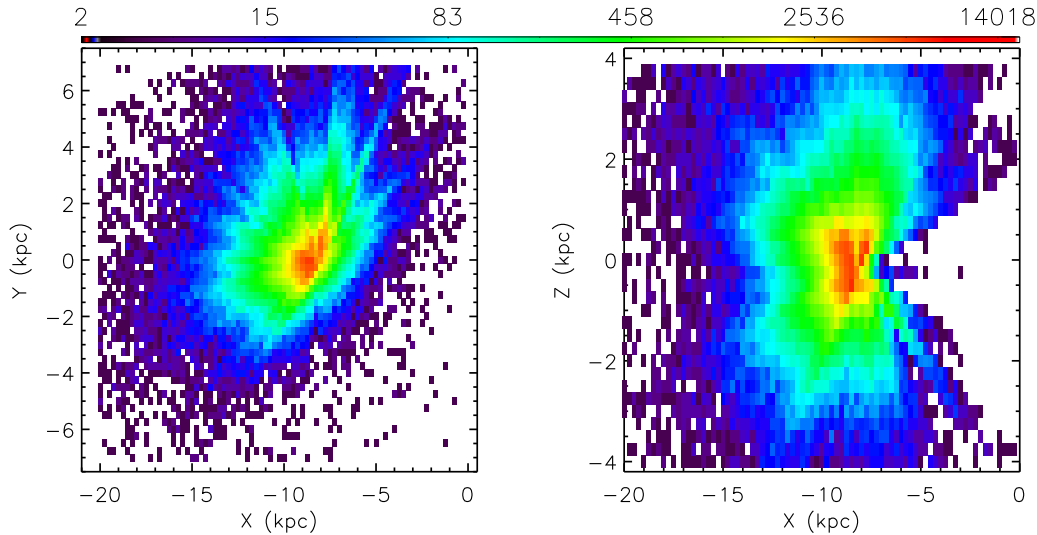}
\caption{Colour-coded stellar density distribution in the Galactic $X$ -- $Y$ plane and $X$ -- $Z$ plane for RGB stars observed with the LAMOST. }
\label{Fig:21}
\end{figure*}

\section{Mass and Age Determination}

\subsection{the Method}

In the last paper of Wu et al. (2018), we investigated the feasibility of deducing mass and age for RGB stars directly from the LAMOST spectra with KPCA-based multivariate linear regression method, taking 4040 stars in the LAMOST-$Kepler$ survey that have accurate seismic parameters from Yu et al. (2018) and Chaplin et al. (2014) as a training data set. We demonstrated that ages thus derived achieve a precison of $\sim$\,24 per cent. For more introduction about the KPCA-based regression method, we refer to the paper of Wu et al. (2018), as well as the concise description in Section\,5.2. One can also refer to Xiang et al. (2017c) for a more detailed example on the application of the method to LAMOST spectra for the determinations of stellar parameters. Here we straightforwardly inherit the method from Wu et al. (2018) but with slightly improvement on the training data set. In Wu et al. (2018), only stars with accurate seismic parameters are adopted as the training sample, and most of them have ${\rm log}\,g<3.3$ so that the parameter space at the higher log\,$g$ is poorly covered. As a consequence, the deduced stellar mass and age for stars near or beyond the log\,$g$ boundary (e.g. base-RGB stars) may suffer large systematics (see their Fig.\,9). To overcome this defect, we expand the parameter coverage of the training sample by adding a number of LAMOST-TGAS sub-giant stars and main-sequence turn-off stars. The supplementary is evenly selected in the $T_{\rm eff}$, log\,$g$ and [Fe/H] space, the sub-giant stars and main-squence turn-off stars are selected with log\ $g$ $>$ 3.8\,dex. Finally, our training sample contains 5376 stars. We plot their distribution in the H-R diagram in Fig.\,8, the effective temperature ranges from 4000\,K to 6600\,K, the surface gravity ranges from 1.5\,dex to 4.5\,dex. Note that there are few stars with log\ $g$ $\sim$ 3.7\,dex and $T_{\rm eff}$ from 5000\,K to 5200\,K because of the selection criteria. However, this small ($\sim$\,0.1\,dex) gap at log\ $g$ $\sim$\,3.7\,dex is not expected to induce significant effect on the determination of mass and age in the current work. Masses and ages of the Gaia-TGAS stars are determined by matching stellar parameters with the $Y^2$ isochrones (Demarque et al. 2004) using the Bayesian method of Xiang et al. (2017b). Here the adopted stellar parameters are $T_{\rm eff}$, [Fe/H] and [$\alpha$/Fe] from the LAMOST spectra and $M_V$ from TGAS parallax. Only stars with $M_V$ error smaller then 0.2\,mag are adopted to quantify the accuracy of the mass and age estimates. Typical uncertainty of the mass estimate is 8 per cent, and that of the age estimates is 23 per cent. By comparing the ages estimated by TGAS-based isochrones and ages estimated by asteroseismology, we find there are about 0.5\,Gyr median difference between TGAS-based isochrone age and asteroseismic age. We have opted to ignore such a small difference since it is not expected to make a major impact for our results. Fig.\,9 plots a color-coded distribution of the (relative) residuals for the regression of age, i.e., the (relative) difference between the spectral age estimated with the KPCA-based regression method and the seismic ages and ages based on isochrones for the training sample. It shows that there are no longer strong systematic patterns across the whole red giant branch, which are essential for obtaining uniform age estimates of the RGB sample stars.

\begin{figure}
\centering
\includegraphics[width=90mm, angle=0]{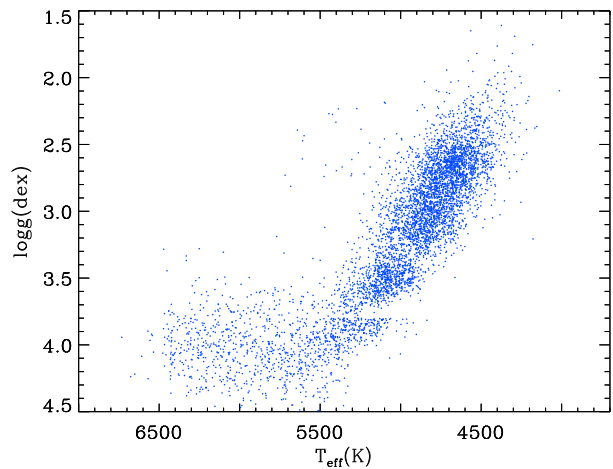}
\caption{Distribution of the 5376 training stars in the $T_{\rm eff}$ -- log\ $g$ plane. }
\label{Fig:10}
\end{figure}

\begin{figure}
\centering
\includegraphics[width=90mm, angle=0]{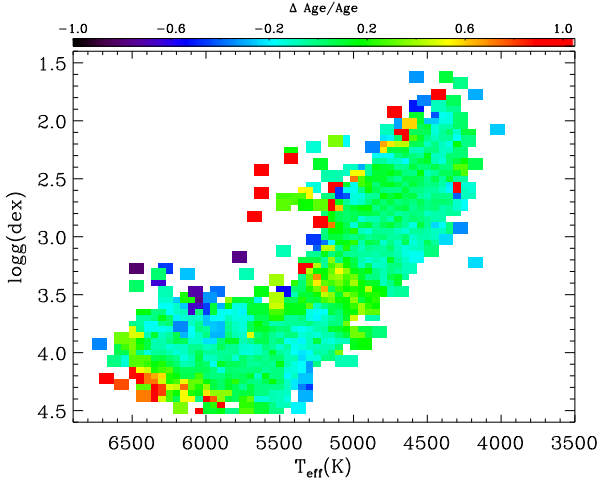}
\caption{Distribution of relative residuals of ages estimated with KPCA method as compared to the seismic values in the $T_{\rm eff}$ -- log\,$g$ plane. }
\label{Fig:11}
\end{figure}
We derive the masses and ages for the whole RGB sample stars. Considering that the KPCA-based multivariate linear regression method are found to be very sensitive to spectral quality (e.g. SNR), and the estimated parameters from low-quality spectra may suffer large systematics (e.g. Xiang et al. 2017c), we implement an internal calibration to the derived ages and masses of the sample stars in the same way as done by Xiang et al. (2017c) for the parameter estimation. In doing so, we first define a parameter $d_{\rm g}$, which represent the maximal kernel value between the target spectrum and any of the spectra in the training set. It is a metric of describing similarities between the target and training spectra. The value of $d_{g}$ will be unity if a target spectrum is exactly the same as one of the training spectra. A small value of $d_{g}$ may be have two possible reasons: one is that the quality of spectra is poor (low SNR). The training set exhibits a relatively uniform S/N distribution above 50, which is the low S/N cut. While as shown in Fig.\,5, the S/N distribution of the full sample exhibits a steep declination from 10 to S/N $>$ 200. The other reason is that the target spectrum is so special that none of training spectra can match it.
Like the method mentioned in the Xiang et al. (2017c), we search for duplicated spectra by requiring one observation of the duplication yields $d_{g}$ $>$ 0.9, for which the derived mass and age are assumed to suffer negligible systematics and can be used as a reference for calibration. Then results derived from spectra that yield $d_{g}$ $<$ 0.9 are grouped into bins of $d_{g}$ with a bin size of 0.05. For each bin, a polynomial model is constructed to calibrate the mass (age) of deduced from spectra with $d_{\rm g} < 0.9$ to that from spectra with $d_{\rm g} > 0.9$.

For the error estimation, we disentangle the error estimation into two parts. One is the random error caused by uncertainty of the spectra, which is a function of spectral SNR, and is deduced by comparing repeat observations that have comparable SNR, as shown in Fig.\, 11. We select duplicate stars that have comparable (differ by less than within 20 per cent) spectral SNR and are collected in different nights, and grouped them into bins of SNR, $T_{\rm eff}$ and [Fe/H]. In each bin, the standard deviation of the differences of parameter values derived from the repeat observations is calculated. The standard deviation, after divided by the square root of 2, is adopted as the random error. The other part of the error estimate comes from the method itself, which is mainly contributed by the age/mass error of the training stars and possible inadequacy of the regression model between age/mass and principal component derived from the spectra with the KPCA algorithm. Here we adopt the standard deviation of the residuals between the regression model predicted (fitted) age/mass and the asteroseismic age/mass and age/mass based on isochrones of the training sample as a measurement of the method error, which is 8 per cent for mass, 26 per cent for age. The estimated error of each stars is the combination of these two parts via $\sqrt{\sigma_{\rm r}^2 + \sigma_{m}^2}$, where $\sigma_{\rm r}$  means random error, and $\sigma_{\rm m}$ represents method error. The distribution of the mass and age estimates as well as their errors are shown in Fig.\,10. For stars with SNR $>$ 30, the median error of mass estimates is 10 per cent, and that of age estimates is 30 per cent. At the top panel of the figure, most stars have masses between 0.8 and 1.8\,$M_{\sun}$. There are few stars with masses larger than 2\,$M_{\sun}$ or smaller than 0.6\,$M_{\sun}$. At the bottom of the figure, the age of stars cover the whole range of possible ages of stars, from close to zero on the young end, up to the age of the universe ($\sim$13.8\,Gyr; Planck collaboration 2016), and the age peaks at 7\,Gyr.

\begin{figure}
\centering
\subfigure{
\label{fig:subfig9:a}
\includegraphics[width=90mm]{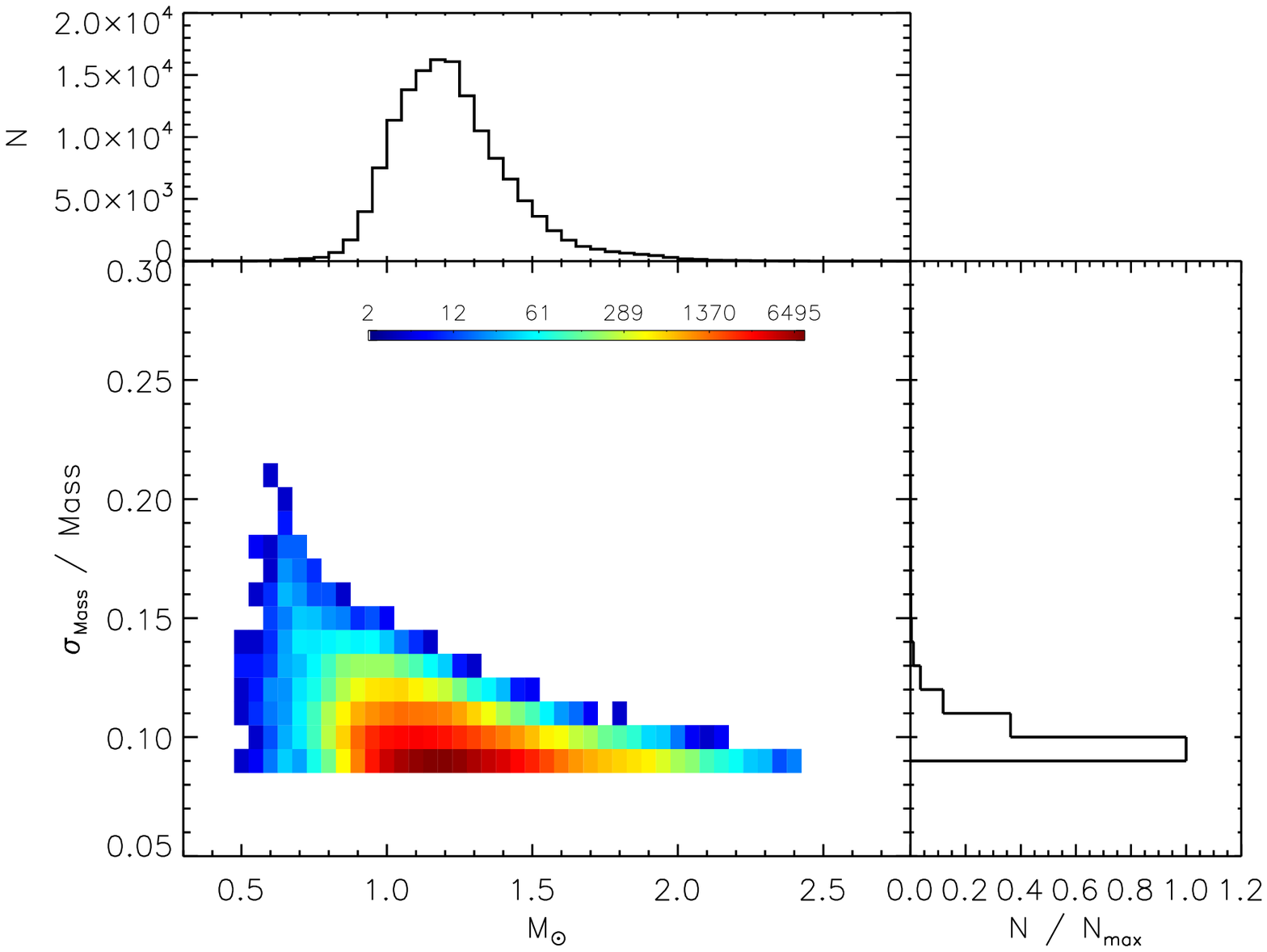}}
\hspace{1in}
\subfigure{
\label{fig:subfig9:b}
\includegraphics[width=90mm]{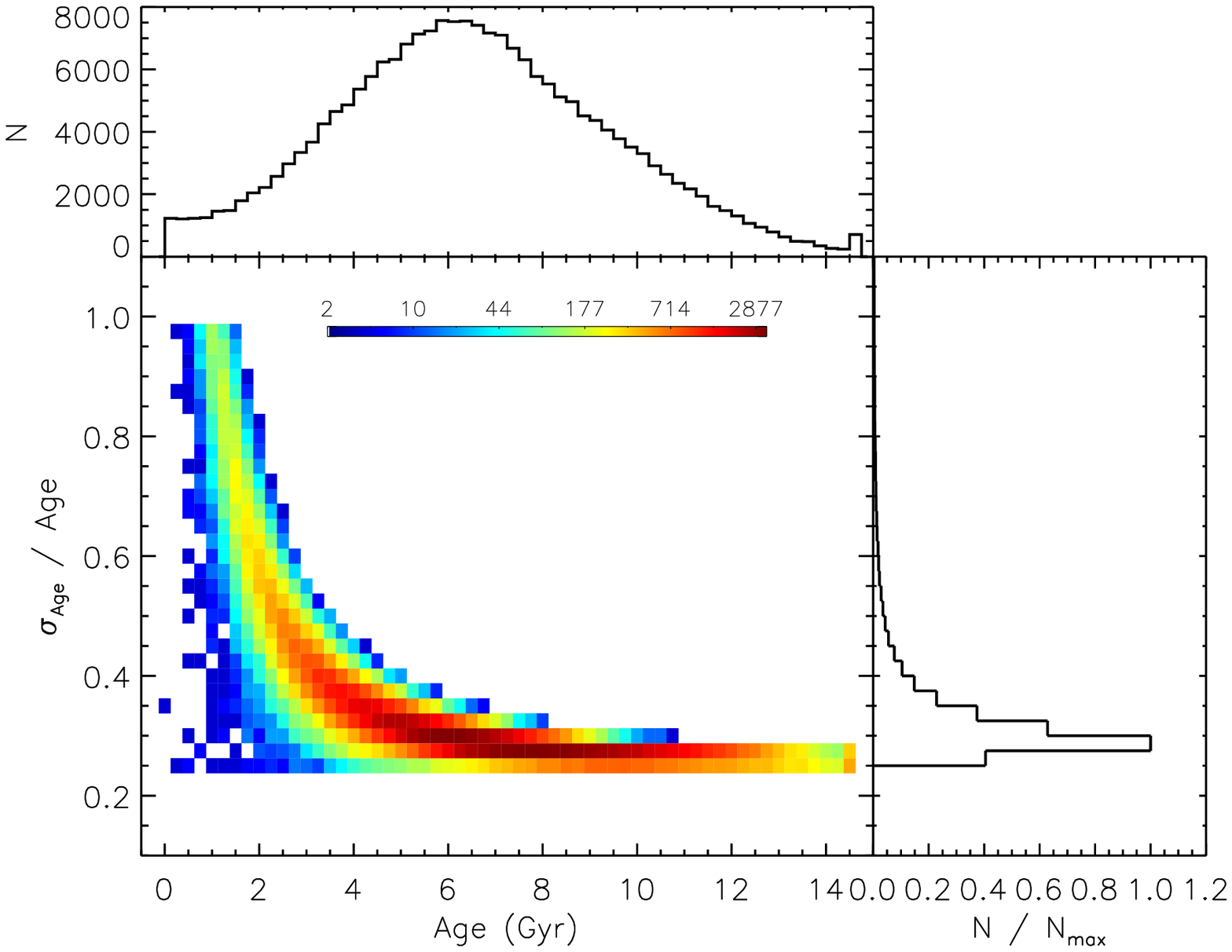}}
\caption{$Upper\,panel$: Distribution of mass estimates, as well as their errors. The upper panel of the figure plots the distribution of mass estimates, the right panel of the figure plots the distribution of their errors of masses. The color bar indicates their density of mass estimates and errors of masses. $Bottom\,panel$: Distribution of age estimates, as well as their errors.}
\label{Fig:subfig9}
\end{figure}

\subsection{Validation of Ages and Masses}

Fig.\,11 plots the internal differences of ages and masses deduced from duplicate observations as a function of the SNR. Here, the duplicate observations refer to those carried out in different nights. The figure shows that the dispersion of ages and masses become nearly flat with the spectral SNR increasing. The relative dispersion of age estimates is 20 per cent when the spectra of SNR is higher than 50, and that of mass estimates is 10 per cent. Therefore, the random error of ages estimated from the spectra is about 20 per cent, and that of masses is about 10 per cent.

\begin{figure}
\centering
\includegraphics[width=90mm, angle=0]{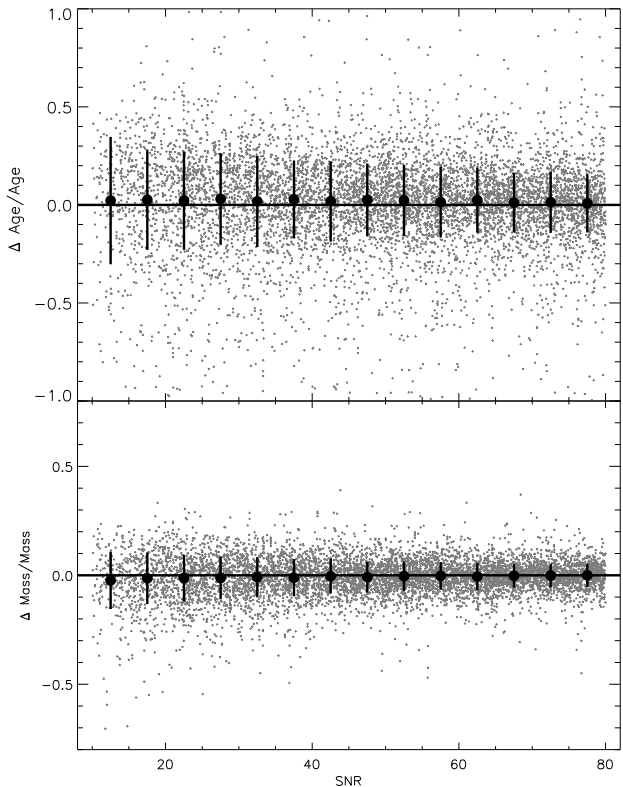}
\caption{Internal differences of age and mass estimates derived from duplicate observations of different spectral SNR, plotted against the SNR of the lower-quality observation. The SNR of the higher quality spectrum is required to be larger than 80 per pixel. The black dots and error bars are respectively the mean and standard deviation for stars in the individual SNR bins. }
\label{Fig:14}
\end{figure}

Open clusters in the Milky Way are generally believed to form from the same gas cloud simultaneously, so that member stars of a cluster have the same age. Ages of cluster members thus provide an independent test for our age estimation. For this purpose, a number of LAMOST plates have been designed to target open clusters of different ages utilizing gray nights reserved for monitoring the instrument performance. Together with data from the main survey, we are able to select RGB stars in three open clusters, namely NGC 2420, M67 (NGC 2682), and Berkeley32. These clusters cover an age range from 2\,Gyr to 6\,Gyr. In the very young and old regimes, we do not find suitable open clusters in the LAMOST plates to test. For younger open clusters, only stars with high mass evolve at RGB state or AGB state, whereas for old open clusters, the number of these open clusters are small in the LAMOST field and usually they are dimmer so that they can hardly observed by LAMOST.

A detailed description of member star identification for these open clusters are presented in the Xiang et al. (2017a). Member stars usually are identified by radial velocities and proper motions. However for some open clusters, if the contaminations of member stars of open clusters from the background stars are severe, besides kinematics parameters for open clusters, we also need distance moduli to discard background stars that deviate significantly ($>$1.5 mag) from the peak values of the clusters. For LAMOST spectra, the accuracy of radial velocities is about 5\,km\,s$^{-1}$, for proper motions, we use the data of UCAC4 \citep{Zacharias13}. From the work of Xiang et al. (2017a), they identify 88, 2858, 72 membership candidates for NGC 2420, M67 and Berkeley 32, respectively. After cross matching with the LAMOST RGB sample stars, there are 9, 43, and 6 members left. The measured ages are obtained by taking the means of individual member stars.

Fig.\,12 presents a direct comparison between the measured cluster ages and the literature values. The figure shows that our ages estimated from the spectra are largely in good agreement with the literature values. For NGC 2420, the dispersion of ages is relatively small, and the number of members in that open cluster is also small. The median age of this cluster is 2.35\,Gyr, and the dispersion is 0.97\,Gyr. It is consistent with the ages given in the literature \citep[2.2\,Gyr;][]{Demarque94,Twarog99}. For M67, The median age is 5.00\,Gyr, with a dispersion of 1.69\,Gyr. The age of this open cluster is overestimated by about 1.00\,Gyr compared with the ages given in the literature \citep[4.0\,Gyr;][]{Demarque92,Carraro94,Dinescu95,Fan96,Richer98,Van04,Schiavon04,Sarajedini09,Barnes16}. The systematic overestimate is probably a consequence of defects of the KPCA-based regression method. For Berkeley 32, the median age of this open cluster is 6.15\,Gyr, with a dispersion of 0.60\,Gyr. The age of this open cluster is also consistent with the ages given in the literature \citep[6.0\,Gyr;][]{Kaluzny91,Richtler01,Salaris04,DOrazi06,Tosi07}.

\begin{figure}
\centering
\includegraphics[width=90mm, angle=0]{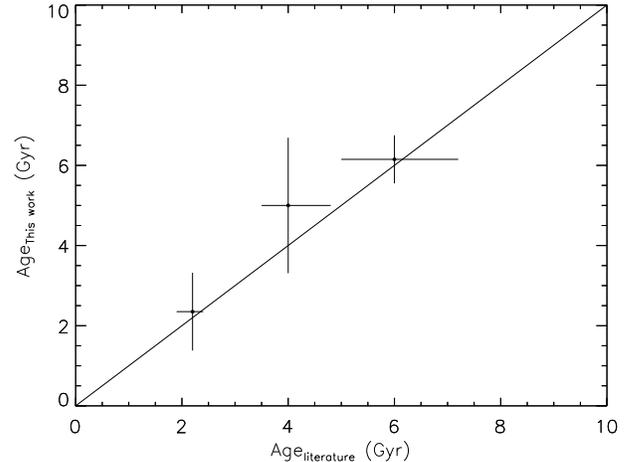}
\caption{Comparison of age estimates of open clusters with literature values. The vertical error bars represent age dispersions (standard deviations) of the individual cluster member stars, while the horizontal error bars represent the range of age estimates in literature. }
\label{Fig:15}
\end{figure}

Sanders \& Das (2018) obtained a catalogue of distances, masses and ages for $\sim$ 3 million stars observed by Gaia with spectroscopic parameters from the large spectroscopic surveys: APOGEE, Gaia-ESO, GALAH, LAMOST, RAVE and SEGUE.
We cross match their catalog with ours and obtained 480,000 common red giant branch stars. After selecting SNR $>$ 50, 2.0\,dex $<$ log\ $g$ $<$ 3.2\,dex, 2000\,K $<$ $T_{\rm eff}$ $<$ 5500\,K, there are 80,000 common red giant branch stars left. Fig.\,13 plots the comparison between the ages estimated by Sanders \& Das (2018) and this work. From the figure, we can see that the ages estimated by Sanders \& Das are generally younger than our work, the ages estimated by them can be younger than 2 -- 3\,Gyr at the older parts. This is because the scaling relation and isochrones we used are different from Sanders \& Das used. We adopt the revised scaling relation (Sharma et al. 2016) when inferring stellar mass from seismic data, while Sanders \& Das (2018) used un-revised scaling relation. In addition, we use the Y$^{2}$ isochrones which consider the effect of $\alpha$-enhanced, whereas Sanders \& Das use the PARSEC isochrones and do not consider [$\alpha$/Fe]. The influence of [$\alpha$/Fe] is vital for older stars. As such, they obtained an age of the thick disk of about 7 Gyr whereas the age from this work is about 10 Gyr.

\begin{figure}
\centering
\includegraphics[width=90mm, angle=0]{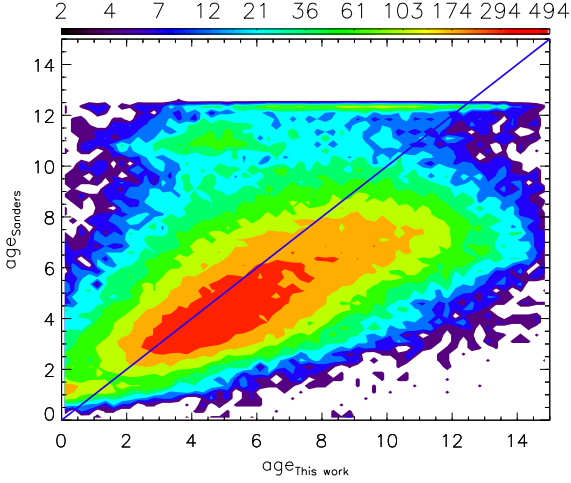}
\caption{Comparison of age estimates of this work with Sanders \& Das (2018). The colour bar represents the numbers of stars. The blue solid line represents the 1:1 line. }
\label{Fig:15}
\end{figure}
\subsection{Age and mass estimates from C, N abundance}

In the last paper of Wu et al. (2018), we also explored the feasibility of estimating ages and masses based on the spectroscopically measured carbon and nitrogen abundances. Utilizing the parameters of Table\,2 and Table\,3 in the Wu et al. (2018), we estimate the ages and masses from the carbon and nitrogen abundances for LAMOST RGB sample stars. We plot the comparison of age estimates from the CN scaling relation with the
KPCA method in Fig.\,14. The figure demonstrates that age estimates with C, N abundances is higher than age estimates from the KPCA method. The median difference is about 0.61\,Gyr, the ages estimated from the CN scaling relation depend on the asteroseismic training data sets, whereas age estimated by KPCA method in this paper added some TGAS sub-giant stars. The ages estimated by TGAS have a median difference of about 0.5\,Gyr compared to the asteroseismic ages. We recommend the ages estimated from the KPCA method rather than the ages estimated from the CN abundances.

\begin{figure}
\centering
\includegraphics[width=90mm, angle=0]{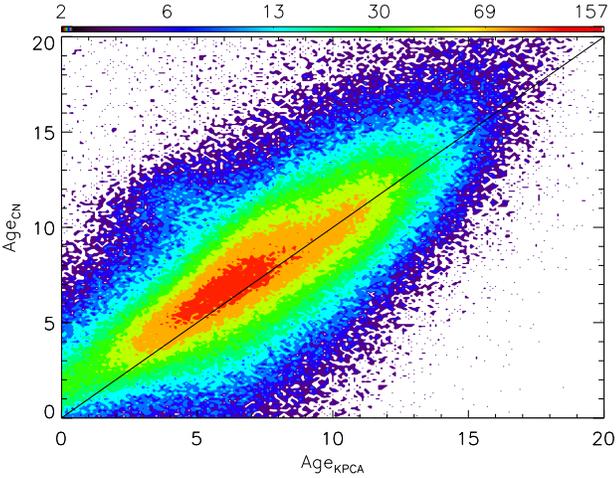}
\caption{Comparison of age estimates between C, N abundance method and KPCA method. The color bar represents the numbers of stars. The black solid line represents 1:1 line. }
\label{Fig:15}
\end{figure}

\section{Relations among age, metallicity and kinematics}

Utilizing this large sample of LAMOST RGB stars, we explore the distribution of stellar ages in the disk $R$ -- $Z$ plane, relations among the age, metallicity and elemental abundances as well as relations between stellar ages and kinematic parameters. To ensure high data quality, stars with SNR lower than 50, surface gravity larger than 3.8\,dex and surface gravity smaller than 2.0\,dex are discarded, leaving about 200\,000 stars in the remaining sample used for the following analysis.

\subsection{Distribution of Stellar Ages in the $R$ -- $Z$ Plane}

Fig.\,15 plots the median age of stars at different positions across the $R$ -- $Z$ plane of the Galactic disk. Here $R$ is the projected Galactocentric distance in the disk midplane, and $Z$ is the height above the disk midplane. Considering that metal-poor stars have large age uncertainties, we discard stars with [Fe/H] $<$ -1.2\,dex. Generally, the data exhibit negative age gradients in the radial direction and positive age gradients in the vertical direction. At small heights, the outer disk of R $\geq9\,$kpc is dominated by young ($\leq$2\,Gyr) stars, which reach larger heights above the disk plane at the farther disk. The inner disk of $R$ $\leq$ 9\,kpc exhibits a positive vertical age gradient for small heights, while the disk at larger heights is dominated by old stars with no significant vertical gradients.

A radial age gradient of the Milky Way was also presented by Xiang et al. (2017a) using main-sequence turnoff and sub-giant stars from the LAMOST Galactic Spectroscopic survey (see their Fig. 23). Our results are clearer than theirs although the overall structures and patterns are similar. The results exhibit strong flaring age structure across the whole radial range from $R\sim6$\,kpc to $\sim13$\,kpc, which is a natural consequence of the disk flaring observed previously via star counts \citep[e.g.][]{Derriere01,Lopez02,Lopez14,Xiang18} and is well reproduced by simulations as a suggested consequence of a weaker restoring force at the outer Galactocentric radii \citep{Minchev15}.

\begin{figure}
\centering
\includegraphics[width=90mm, angle=0]{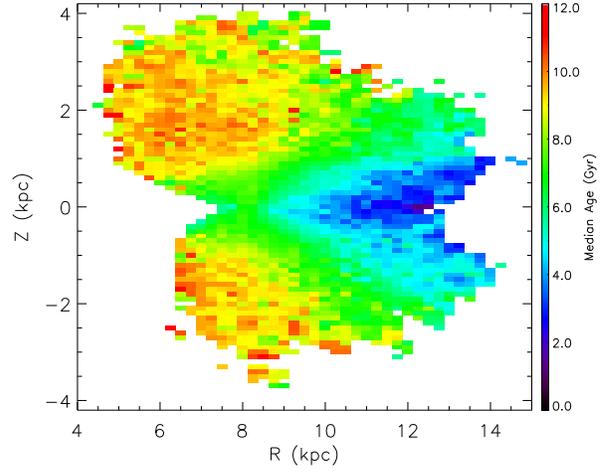}
\caption{Color-coded distribution of the median age for stars of different spatial bins in the R -- Z plane. The adopted bin size is 0.25\,kpc in the R direction and 0.1\,kpc in the Z direction. }
\label{Fig:20}
\end{figure}

Fig.\,16 plots the $R$ -- Age relation in different bins of $|Z|$. We divided the sample into four vertical bins: 0\,kpc $< |Z| <$ 0.5\,kpc, 0.5\,kpc $< |Z| <$ 1.0\,kpc, 1.0\,kpc $< |Z| <$ 1.5\,kpc and 1.5\,kpc $< |Z| <$ 2.0\,kpc, and plot the median age as a function of $R$ for each vertical bins of $|Z|$. At solar radius, we find a median age for RGB stars of about 7\,Gyr in the midplane, increasing to 9.5\,Gyr for 1.5\,kpc $< |Z| <$ 2.0\,kpc. This is slightly underestimated compared to the vertical age gradient of 4 $\pm$ 2 Gyr kpc$^{-1}$ at the solar radius by Casagrande et al. (2016). It can be found that at any given radius, the median age of RGB stars increases with height above the disk. We also find radial age gradients at almost all heights above the midpline. At about 1 -- 2\,kpc above the disk, stellar ages go from about 9 -- 10\,Gyr in the inner disk to about 5\,Gyr in the outer disk. The results is similar with the study of radial age gradient of the geometrically thick disk in the Martig et al. (2016).

\begin{figure}
\centering
\includegraphics[width=90mm, angle=0]{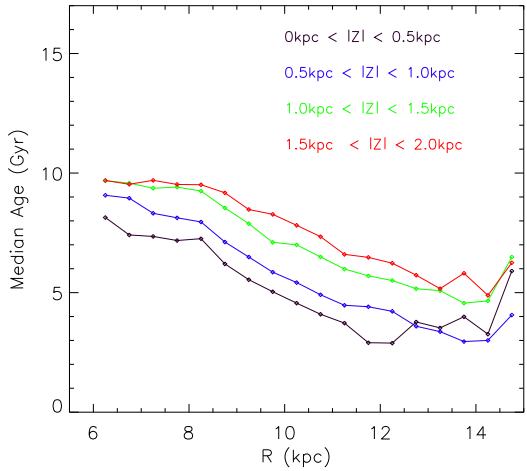}
\caption{R -- Age relation in different bins of $|Z|$. The color represent different bins of $|Z|$, the black lines mean $|Z|$ = 0\,kpc, the blue lines mean $|Z|$ = 0.5\,kpc, the green lines mean $|Z|$=1.0\,kpc, and the red lines mean $|Z|$ = 2.0\,kpc.}
\label{Fig:20}
\end{figure}

\subsection{Age, [$\alpha$/Fe] and [Fe/H] Relation}
Fig.\,17 plots the distribution of the median stellar ages and masses in the individual mono-abundance bins of the [Fe/${\rm H}$] -- [$\alpha$/Fe] plane. To show the potential patterns better, only stars with ${\rm SNR} > 50$, $2 < {\rm log}\,g < 3.8$, $T_{\rm eff} < 5500$\,K and ${\rm [Fe/H]} > -1.2$\,dex are used to generate the figure because age estimates for stars in these parameter ranges have higher accuracy. The figure shows clear patterns in the distribution of median stellar ages and masses across the [Fe/${\rm H}$] -- [$\alpha$/Fe] plane. For a given [Fe/${\rm H}$], stars of higher [$\alpha$/Fe] have older ages. There is an old ($>10$\,Gyr) sequence of stars on the high-$\alpha$ side. We could also divide them into three region of ages, including stars with ages older than 10\,Gyr, intermediate-age and young. For old stars, their metallicity is generally distributed between -0.7\,dex and 0.3\,dex, with [$\alpha$/Fe] higher than 0.10\,dex; for intermediate-age stars, their [$\alpha$/Fe] value is distributed from 0\,dex to 0.3\,dex, the low [$\alpha$/Fe] region is dominated by young stars. Note that, the current results deviate to some extent from those of Xiang et al. (2017a) and Haywood et al. (2013), and also different from our former results from the asteroseismic sample (Wu et al. 2018) in the metal-poor side (${\rm [Fe/H]}\lesssim-0.6$\,dex), as the current results suggest that those metal-poor, $\alpha$-enhanced stars are dominated by intermediate-age stars whereas previous studies suggested they are old stars. We believe this is because a defect of the multivariate linear regression method. Since the regression is carried out in a rather high dimensional space (100), it is easily biased near the boundary of the parameter space where the number of training stars is small.
The ages and masses of metal-poor stars should be carefully used for these metal-poor stars.
\begin{figure}
\centering
\includegraphics[width=100mm, angle=0]{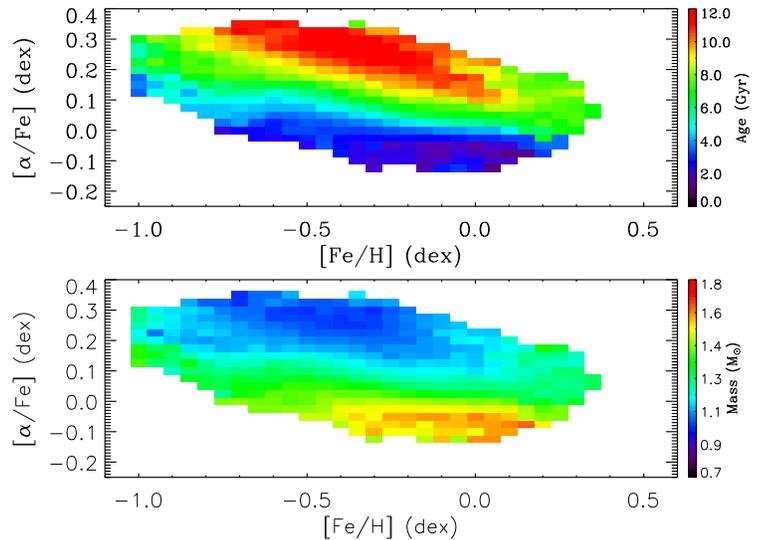}
\caption{$Upper panel$: Distribution of median stellar ages of [$\alpha$/Fe] -- [Fe/${\rm H}$]. $Lower panel$: Distribution of median stellar masses of [$\alpha$/Fe] -- [Fe/${\rm H}$]. The color bar represents the median mass and age of each bin. }
\label{Fig:16}
\end{figure}

Fig.\,18 plots the stellar number density distribution in the [Fe/H] -- [$\alpha$/Fe] plane for stars in different age bins. Here we use the same RGB stellar sample mentioned above. The figure shows that for all individual age bins, stars exhibit wide distribution in the [Fe/H] -- [$\alpha$/Fe] plane, implying that in a given mono-abundance bin of [Fe/H] and [$\alpha$/Fe], stars could have an extensive age distribution, especially for bins of intermediate abundances (e.g., -0.5\,dex $\leq$ [Fe/H] $\leq$ 0\,dex, 0\,dex $\leq$ [$\alpha$/Fe]$\leq$ 0.1\,dex). Nevertheless, the figure demonstrates a clear temporal evolution trend of [Fe/H] -- [$\alpha$/Fe] sequences. Both the low-$\alpha$ and high-$\alpha$ sequences are presented in the age bin of 8 -- 10\,Gyr. As the age increases from 0 -- 2\,Gyr to 8 -- 10\,Gyr, [$\alpha$/Fe] values of the lower-$\alpha$ sequence at solar metallicity increase from about -0.1\,dex to about 0.0\,dex.
Such a double-sequence feature is consistent with the widely suggested thin and thick disk sequences \citep[e.g.][]{Fuhrmann98,Bensby03,Lee11,Haywood13,Hayden15,Xiang17b}.
\begin{figure*}
\centering
\includegraphics[width=150mm, angle=0]{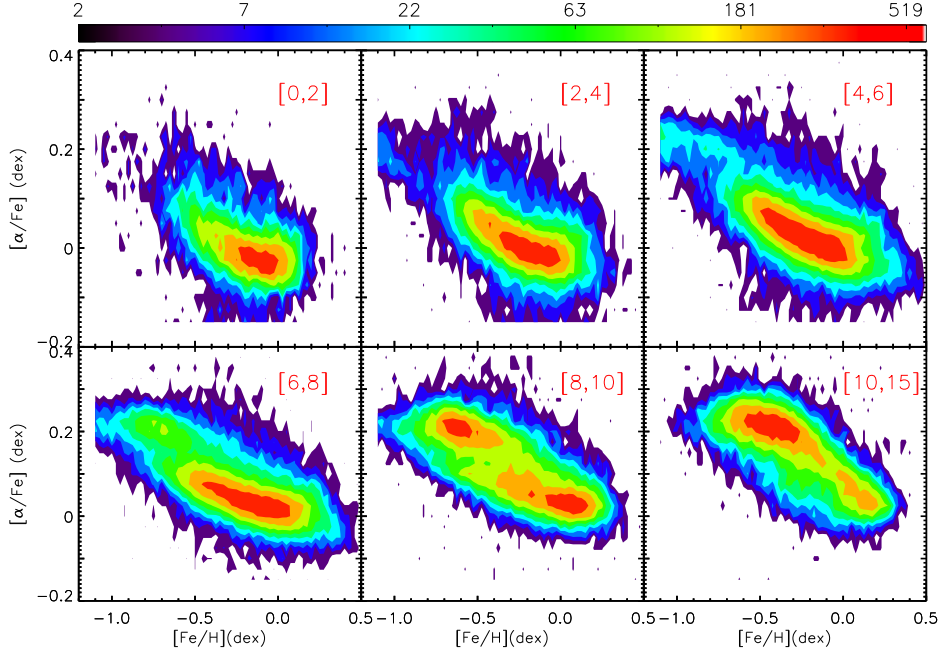}
\caption{Colour-coded stellar number density in the [Fe/H] -- [$\alpha$/Fe] plane for stars in different age bins, as labeled in red on the upper-right corner of the figures. The bin size is 0.05\,dex in [Fe/H], and 0.025\,dex in [$\alpha$/Fe].}
\label{Fig:19}
\end{figure*}

Fig.\,19 plots the density distribution of RGB stars in the age -- [$\alpha$/Fe] plane, the color bar represents the number of stars in logarithmic scale. The figure exhibits two prominent sequences of different [$\alpha$/Fe] values. Stars younger than 8\,Gyr belong to a sequence of lower [$\alpha$/Fe] value, and the [$\alpha$/Fe] slowly increases with the median age of RGB stars in an approximately linealy manner with a slope of $\leq$ 0.02\,dex/Gyr. At the older end, the high-$\alpha$ sequence, which has an almost constant [$\alpha$/Fe] value about 0.2\,dex for stars older than 10\,Gyr. Note that, the current sample also include a number of young ($<$ 6\,Gyr) stars with unexpectedly high [$\alpha$/Fe] values and old ($>$ 10\,Gyr) stars with low [$\alpha$/Fe] values. Such young, high-$\alpha$ stars are also found in some previous works \citep[e.g.][]{Chiappini15,Martig15,Wu18}. For these stars, one possible explanation is that they were formed near the ends of the Galactic bar or may be evolved blue stragglers. For old, low-$\alpha$ stars, they may be formed at the inner disk. At the younger end, the high-$\alpha$ sequence is consistent with the results from Figure 20. The presence of two age -- [$\alpha$/Fe] sequences either suggests the existence of two distinct phases of formation history of the Galactic disk \citep[e.g.][]{Haywood13,Xiang15a,Xiang17a,Grisoni18,Spitoni18} or is a natural consequence of a continuous disk formation process \citep[e.g.][]{Schonrich09}.

\begin{figure}
\centering
\includegraphics[width=90mm, angle=0]{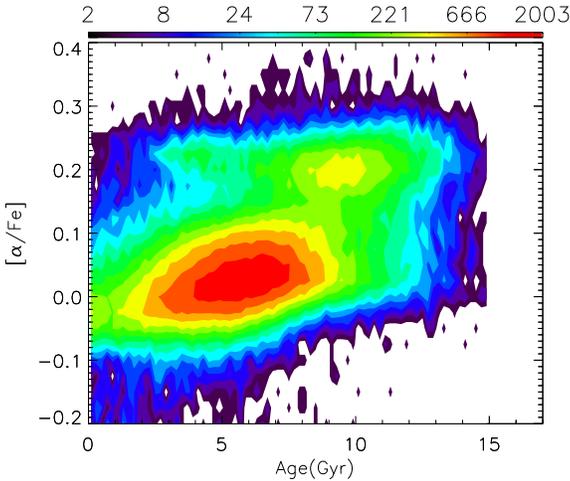}
\caption{Age -- [$\alpha$/Fe] relation for the RGB sample stars, the color bar represents the number of stars in bins of 0.2\,Gyr $\times$ 0.025 (bin size) dex. }
\label{Fig:17}
\end{figure}

Fig.\,20 plots the density distribution of the RGB stars in the age --[Fe/H] plane. The figure shows that stars exhibit a wide range of [Fe/${\rm H}$] values for all ages. For stars older than 8\,Gyr, there is a trend that age increases with decreasing [Fe/${\rm H}$] but there are still quite a large fraction of old metal-rich stars which may be born in the inner disk. The broad range of [Fe/${\rm H}$] values for disk stars at a given age may suggest a complicated disk chemical and dynamic history. As the sample stars cover a large volume, one possible cause of the broad [Fe/${\rm H}$] distribution is the existence of both radial and vertical [Fe/${\rm H}$] gradients for mono-age stellar populations. However, it is also found that even in a limited volume, for instance, the solar neighborhood, the age -- [Fe/${\rm H}$] relation for young stars still exhibit a broad distribution. The inevitable presence of mixing of stars born at different positions caused by stellar radial migration \citep[e.g.][]{Sellwood02,Roskar08,Schonrich09,Loebman11} has certainly played an important role for such [Fe/${\rm H}$] broadening. The results are similar with previous studies \citep[e.g.][]{Haywood13,Xiang17a}. 
However the current sample exhibits a larger fraction of old metal-rich stars which should studied further.

\begin{figure}
\centering
\includegraphics[width=90mm, angle=0]{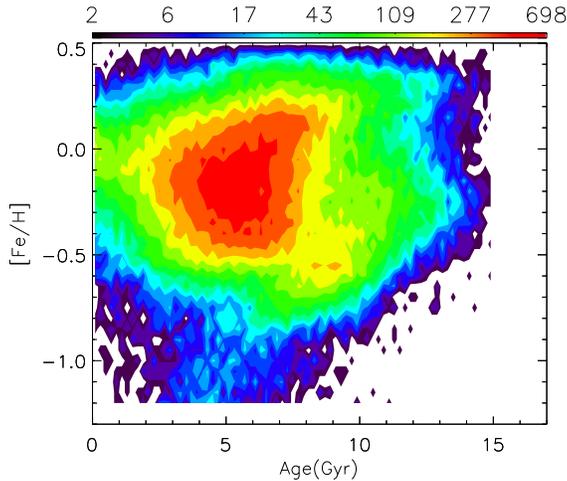}
\caption{Age -- [Fe/H] relation for the RGB sample stars, the color bar represents the number of stars in bins of 0.2\,Gyr $\times$ 0.05 (bin size) dex.  }
\label{Fig:18}
\end{figure}

\subsection{Stellar age versus kinematics parameters}

The orbital properties of stars in the disk carry signatures of their birth place, but they are also expected to change over time due to the dynamical (secular) evolution of the Galaxy. An important process in secular evolution is radial mixing \citep[e.g.][]{Sellwood02,Roskar08,Schonrich09,Loebman11,Loebman16,Minchev14}, in which stars are redistributed through either changes in angular momentum (churning) or non-circular orbital motion (blurring), which alters the radius of their orbit.

Stellar orbits can be quantified by two dynamical parameters: orbital eccentricity and orbital (vertical) angular momentum which can be derived with precise measurements of distance and velocity. We cross-match the LAMOST RGB sample stars with Gaia DR2 and compute their orbits using the axisymmetric and steady Galactic potential of Gardner \& Flynn (2010). The distance is from the inverse of parallax ($\varpi$) from Gaia DR2. We correct systematic difference of the $\varpi$ with 0.029$^{''}$ \citep{Gaia18a}. Note that distance inferred from the inverse of the parallax becomes biased when the uncertainty in parallax is larger or the parallax is small, and one should use the distances and kinematic parameters stars with large parallax errors with cautions. In our sample, 82 per cent of the RGB stars have parallax errors smaller than 20 per cent.

We first examine the two orbital parameters versus age in Fig.\,21. The left panel of the figure shows eccentricity as a function of stellar age, the color bar represents the logarithm number of stars. Generally, it shows that most of the red giant stars have eccentricities ranging from 0 to 0.4. Both the median value and the dispersion of the eccentricity increase with increasing age. Over the 2 -- 12\,Gyr timescale, the typical eccentricity increases from 0.1 to 0.3, corresponding to an increasing rate of 0.02 per Gyr. The dispersion of the eccentricity also increases with age, from 0.07 at 2\,Gyr to 0.14 at 10\,Gyr. In the right panel of the figure shows the angular momentum ($L$) as a function of age. Conversely to the eccentricity, the angular momentum fairly smoothly decreases from 2 to 12\,Gyr. with a gradient of about -50\,kpc\,km$^{-1}$\,s$^{-1}$\,Gyr$^{-1}$.

\begin{figure*}
\centering
\includegraphics[width=180mm, angle=0]{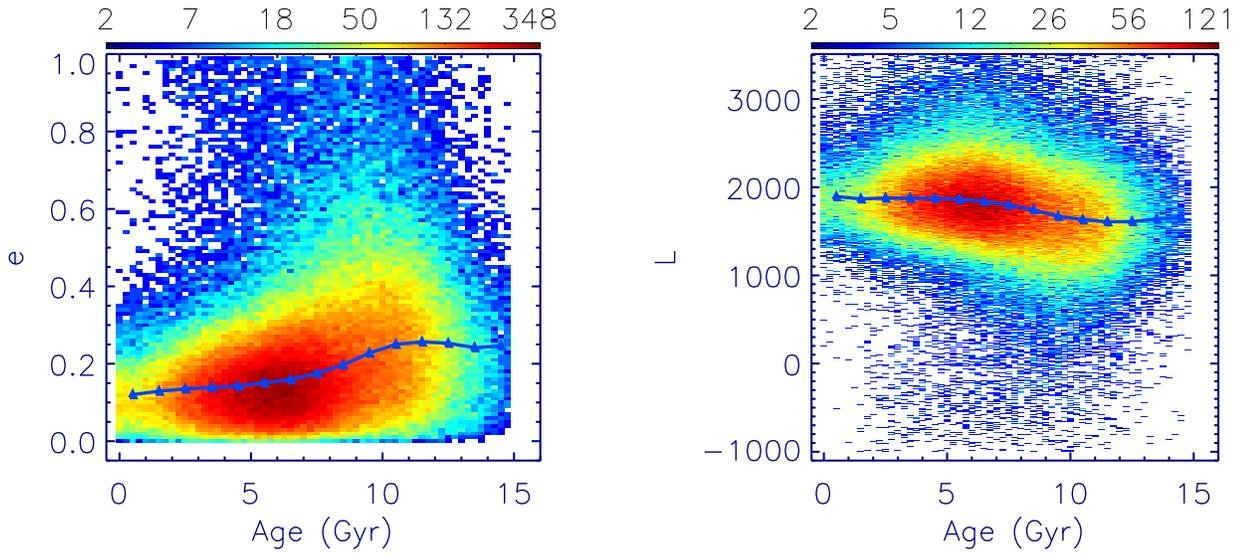}
\caption{$Left panel$: distribution of stars in the age -- e plane, the color bar represents the logarithm of the number of stars. $Right panel$: distribution of stars in the age -- L plane, the color bar represents the logarithm of the number of stars. }
\label{Fig:18}
\end{figure*}

Fig.\,22 plots the distribution of the RGB stars in the age -- $\upsilon_{R}$, age -- $\upsilon_{\varphi}$, and age -- $\upsilon_{Z}$ plane. Note that the motion of the Sun with respect to LSR was corrected using the suggested values of Huang et al. (2016). In the left panel of the Figure, we can find that most of the RGB stars are located in the -100 to 100\,km s$^{-1}$ for $\upsilon_{R}$. The median value of $\upsilon_{R}$ is 7\,km s$^{-1}$, and the dispersion is 47\,km s$^{-1}$. The dispersion becomes larger for older stars.
At the middle of the Figure, it can be found that most of the RGB stars locate in the range of 100 to 300 km s$^{-1}$ for $\upsilon_{\varphi}$. The median value and the dispersion for $\upsilon_{\varphi}$ are 200\,km s$^{-1}$ and 34\,km s$^{-1}$, respectively. The velocity becomes lower as the age increases. As we discussed in Section 4.2, the age estimates of metal-poor stars may have been underestimated, and this causes the velocity dispersion for stars of 8 -- 12\,Gyr looking to be larger than that of the oldest stars ($>$12\,Gyr). However, a quantitative computation shows that the values of velocity dispersion for the 8 -- 12\,Gyr and the $>$ 12\,Gyr population are comparable, which is because the number of metal-poor stars in our sample is small compared to the metal-rich stars.

 In the right panel of the Figure, we can see that most of the RGB stars are located in the -50 to 50 km s$^{-1}$ for $\upsilon_{Z}$, the shape is similar with age -- $\upsilon_{R}$ plane. For $\upsilon_{Z}$, the median value and the dispersion are -2\,km s$^{-1}$ and 27\,km s$^{-1}$.

\begin{figure*}
\centering
\includegraphics[width=180mm, angle=0]{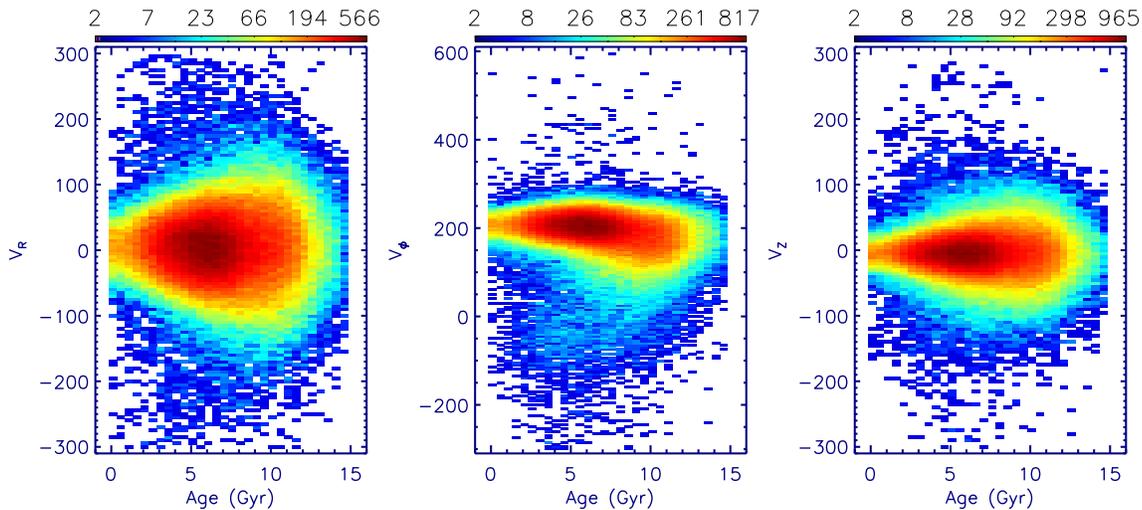}
\caption{$Left panel$: distribution of stars in the age -- $\upsilon_{R}$ plane, the color bar represents the logarithm of the number of stars. $Middle panel$: distribution of stars in the age -- $\upsilon_{\varphi}$ plane, the color bar represents the logarithm of the number of stars. $Right panel$: distribution of stars in the age -- $\upsilon_{Z}$ plane, the color bar represents the logarithm of the number of stars. }
\label{Fig:18}
\end{figure*}

\section{Conclusion}

In this paper, we present a catalog of mass and age estimates for 640\,986 RGB stars from the LAMOST Galactic Survey. The RGB stars are identified with log\,$g$ and $\Delta P$ derived from the LAMOST spectra. Examinations suggest that for stars with SNR $>$ 50, which occupy 31 per cent of the whole RGB sample, contaminations from RC stars is only 2 per cent. Mass and age of the RGB stars are derived with a data-driven method based on KPCA-based multivariate linear regression, utilizing stars with asteroseismic masses and ages as well as sub-giants with isochrone ages as the training data set. Typical error of age estimates is 30 per cent, and that of mass estimates is 10 per cent for stars with SNR higher than 30. Note that although the age estimates for the overall sample stars are robust, ages of the metal-poor stars (${\rm [Fe/H]}\lesssim0.5$) may suffer non-negligible bias due to inadequacy of the current method. In the catalog, we also provide masses and ages derived from their C and N abundances.

With this catalog, we have investigated the stellar age distribution in the disk $R$--$Z$ plane and in the [Fe/H]--[$\alpha$/Fe] plane, the stellar density distribution in the [Fe/H]--[$\alpha$/Fe] plane for mono-age populations, the age--[$\alpha$/Fe] and age--[Fe/H] relation. We have also explored the age -- velocity relation, the age -- eccentricity and age -- vertical angular momentum relation utitlizing stellar orbital parameters derived with the LAMOST and Gaia DR2 data. We find that:

(i)the RGB stars exhibit negative age gradients in the radial and positive age gradients in the vertical direction. The age structure of the disk is significantly flared across the whole disk of $6<R<13$\,kpc. At small heights, the outer disk of R $>$ 9\,kpc is dominated by young ($<$ 2\,Gyr) stars, which reach larger heights above the disk plane at the outer disk. At solar radius, the median age of the RGB stars is about 7\,Gyr in the midplane, increasing to 9.5\,Gyr at 1.5\,kpc $< |Z| <$ 2.0\,kpc.

(ii)There are clear patterns in the distribution of median stellar ages across the [Fe/H] -- [$\alpha$/Fe] plane. For a given [Fe/H], stars of higher [$\alpha$/Fe] have older ages. There is an old ($>$ 10\,Gyr) sequence of stars on the high-$\alpha$ side. For all individual age bins, stars exhibit wide distribution in the [Fe/H] -- [$\alpha$/Fe] plane, implying that in a given mono-abundance bin of [Fe/H] and [$\alpha$/Fe], stars could have an extensive age distribution, especially for bins of intermediate abundances (e.g., $-$0.5\,dex $<$ [Fe/H] $<$ 0\,dex, 0\,dex $<$ [$\alpha$/Fe] $<$ 0.1\,dex). Stellar distribution in the age -- [$\alpha$/Fe] plane exhibits two prominent sequences of different [$\alpha$/Fe] values: a low-[$\alpha$/Fe] sequence which contains stars from very young age to older than 10\,Gyr, and the [$\alpha$/Fe] slowly increases with the age in an approximately linearly manner with a slope of $\sim$ 0.02\,dex\,Gyr$^{-1}$. The high-$\alpha$ sequence has an almost constant [$\alpha$/Fe] value about 0.2 dex and it contains stars from $<5$\,Gyr to $>12$\,Gyr, with a typical (median) age of 10\,Gyr. Our results show a significant fraction of old, metal-rich (${\rm [Fe/H]}\gtrsim$0) stars. They are probably migrators from the inner disk. .

(iii) The orbital eccentricity of the RGB sample stars increase with age with a rate of about 0.02\,Gyr$^{-1}$. The dispersion of the orbit eccentricity also increases with age. The angular momentum fairly smoothly decreases with age from 2 to 12\,Gyr, with a slope of about -50\,kpc\,km$^{-1}$\,s$^{-1}$\,Gyr$^{-1}$.

\vspace{7mm} \noindent {\bf Acknowledgments}
{ This work is supported by the Joint Research Fund in Astronomy (U1631236) under cooperative agreement between the National Natural Science Foundation of China (NSFC) and Chinese Academy of Sciences (CAS), the National Key Basic Research Program of China 2014CB845700 and Joint Funds of the National Natural Science Foundation of China (Grant No. U1531244) and the National Natural Science Foundation of China (NSFC) under grant 11390371. M.-S. Xiang, Y. Huang and C. Wang acknowledge supports from NSFC Grant No. 11703035.

Guoshoujing Telescope (the Large Sky Area Multi-Object Fiber Spectroscopic Telescope LAMOST)
is a National Major Scientific Project built by the Chinese Academy of Sciences.
Funding for the project has been provided by the National Development and Reform Commission.
LAMOST is operated and managed by the National Astronomical Observatories, Chinese Academy of Sciences.}

{}

\label{lastpage}

\end{document}